\documentclass[epj]{svjour}
\usepackage{graphics}
\usepackage{cite}

\begin{document}
\title{Identification of Jet-like events using a Multiplicity Detector}
\author{Ranbir Singh and  Bedangadas Mohanty}                     % Do not remove
%
%\offprints{*email: ransinghep@gmail.com}          % Insert a name or remove this line
%

\mail {ransinghep@gmail.com}

\institute{School of Physical Sciences, National Institute of
   Science Education and Research, HBNI, Jatni-752050, India}
\date{}

\abstract{
We present a method for studying the detection of jets in high
energy hadronic collisions using multiplicity detector in forward
rapidities. Such a study enhances the physics scope of multiplicity 
detectors at forward rapidities in LHC.  At LHC energies the jets may be produced
with significant cross section in forward rapidities. A multi
resolution wavelet analysis technique can locate the spatial position
of jets due to its feature of space-scale locality. The discrete
wavelet proves to be very effective in probing physics simultaneously
at different locations in phase space and at
different scales to identify jet-like events. The key feature this
analysis exploits is the difference in particle density in localized
regions of the detector due to jet-like and underlying events. We find that this method has a significant
sensitivity towards detecting jet position and its size.  The jets can
be found with the efficiency and purity of the order of 46\%.
%
%\PACS{
%      {PACS-key}{discribing text of that key}   \and
%      {PACS-key}{discribing text of that key}
%     } % end of PACS codes
} %end of abstract
\maketitle
\section{Introduction}
\noindent At Large Hadron Collider (LHC), with the large center-of-mass (CM)
energy, multi-jet events may be produced with measurable cross-section
in forward rapidities~\cite{D0}. Typical 3-jets events arising from $qg
\rightarrow qgg$ : $gg \rightarrow gg$ should appear in the ratio of
0.3:1 as  discussed in~\cite{3jets}. If the CM system has a boost either in +ve
or in the -ve z-direction, the jets might be directed in the forward
rapidity.  For the partonic interaction of a low ``x'' gluon and the
high  ``x''  quark will lead to a jet in forward direction, where x
is  the fraction of momentum carried by the partons.
One of the first measurement of inclusive jet production cross section
in forward rapidities is performed in $p\bar p$ collision at $\sqrt{s}$ = 1.8 TeV
with D0 detector at the Fermilab Tevatron~\cite{D0}. The differential cross-
section $d^{2}\sigma/(dE_{T} d\eta$) was measured upto $|\eta|$ $<$ 3,
where $E_{T}$ is the transverse energy of the jet, $\sigma$ is the
cross-section for jet production and $\eta$
is the jet pseudorapidity. The results are found to be in good
 agreement with next-to-leading order predictions from QCD~\cite{nlo_jets} and
indicate a preference for certain parton distribution functions.

In pp collisions, dijet events will appear with jets lying back to
back in  azimuthal angle. In high energy experiments e.g. ALICE 
 these may be easily studied  using the central barrel detectors.
 However one may encounter
events  where the barrel detectors see two of more than two jets in an event where
the  toplogy may suggest a missing jet which may be in other part of
the  phase space (forward rapidity). If even the direction of such a missing jet can be
found,  more physics can be extracted from such an event.
Jets in general produce particles which are confined to a cone and
hence the  spatial particle density within the jet region is expected
to be very  diferent compared to a normal event in pp collisions. 

The aim  of this study is to explore if this disticntion of localized
high particle multilicity density can be
exploited  successfully to predict the jet direction or identify
jet-like events. This study is aimed on the present high energy
experiments STAR (at RHIC) and ALICE (at LHC). In the forward region
of such experiments there are a set of
charged particle detectors and a photon multiplicity detector. 
We have conducted this study using charged particle in the forward
rapidity covering 2.3 $<$ $\eta$ $<$ 3.9.

 In this paper, we have used a
multiresolution analysis by discrete wavelet transformation (DWT)
which has been successfully used in engineering, mathematics and
computer science, astrophysics and multiparticle
productions, DCC search~\cite{Fang:1996ju, Huang:1995tv, Randrup:1997kt, wavelet1, wavelet2, wavelet3, wavelet4, wavelet5}. 
In our study, we demonstrate
that the DWT  proves to be very useful in identifying jet-like events
in terms of their position and size. There is no information loss due
to completeness and orthogonality of the DWT basis.
Also the wavelet transformation has advantages over traditional
 Fourier methods in analyzing physical situations where the signal
 contains discontinuities and sharp spikes. We have applied the
wavelet transformation to the data of the PYTHIA~\cite{pythia} Monte
Carlo simulation generated for p-p collisions at $\sqrt{s}$ = 7 TeV.

\section{Model description}
The simulation study is carried out using PYTHIA6~\cite{pythia}
  event generator
version 6.4. To make the results more realistic, we have applied a
detetor response of such a multiplicity detector with typical values of efficiency
and purity 60\%. We have generated two sets of data, one for ``minimum
bias''  and other for ``jets'' in pp collisions at $\sqrt{s}$ = 7
TeV. Since we are using charged particles in our study, the magnetic
field could have significant effect and thus the study is also
extended by incorporating field ON and OFF conditions. 

\begin{figure}
     \resizebox{0.5\textwidth}{!}{\includegraphics*{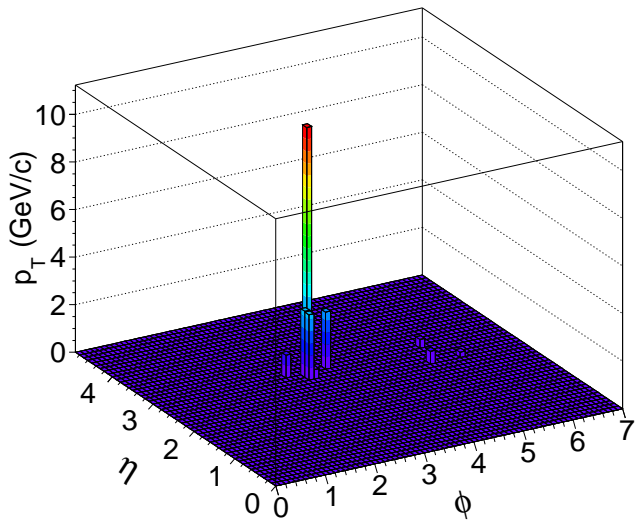}}
     \caption{\label{fig:lego} A typical PYTHIA simulated jet event of jet $E_{T}$ $>$
   20 GeV falling on forward rapidity (2 $<$ $\eta$ $<$ 4)
   in p-p collisions at $\sqrt{s}$ = 7 TeV.}
   \end{figure}

\begin{figure}
     \resizebox{0.45\textwidth}{!}{\includegraphics*{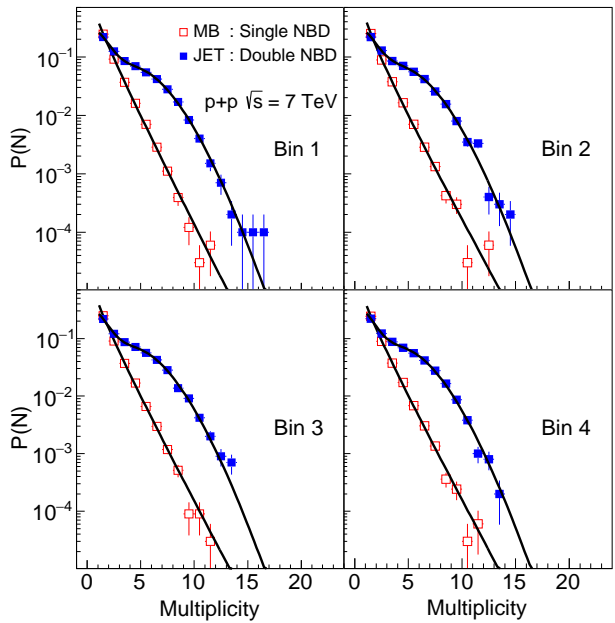}}
     \caption{\label{fig:Chap6_mult7tev} Multiplicity distribution for
  minimum bias (red marker) and jet events (blue marker) in
  four particular $\eta-\phi$ bins in p-p collisions at $\sqrt{s}$ =
  7 TeV from MC truth.}
   \end{figure}

The di-jet events are generated without any initial and final state
radiations and at least one of the jets is allowed to fall in the
forward rapidities (2.3 $<$ $\eta$ $<$ 3.9). The jet transverse energy is
taken as $E_{T}$ $>$ 20 GeV . A typical distribution in $\eta$ - $\phi$
plane of particles  from a jet event after putting a transverse
momentum ($p_{T}$) threshold
of 2 GeV/c on the produced particles is shown in Fig.~\ref{fig:lego}. 
Fig.~\ref{fig:Chap6_mult7tev} shows particle density distribution for
minimum bias (MB) and jet events in $\eta$ coverage, 2.3 $< \eta <$
3.9 for four particular $\eta$ -  $\phi$ bins at $\sqrt{s}$ = 7
TeV. These bins are the one where we allowed the jet to fall on. 
The multiplicity distribution for jets is fitted with double Negative
 Binomial Distribution (NBD) function and that for minimum bias is 
single NBD fit. The double NBD is needed to account for two types 
of particle production processes, one due to non-jet (soft processes)
 and other due to jets (hard processes).

Throughout this paper, the generated level result are named as MC truth
and after applying detector response as Digits.

\section{A multi-resolution discrete wavelet technique}
A discrete wavelet transform (DWT) is any wavelet transform for which
the wavelets are discretely sampled. The DWT 
generalizes the standard Fourier analysis. We have performed our study
with the idea of multiresolution analysis by using the well known Haar
wavelet. For an input represented by a list of $2^n$ numbers, the Haar
wavelet transform may be considered to simply pair up input values,
storing the difference and passing the sum. This process is repeated
recursively, pairing up the sums to provide the next scale: finally
resulting in $2^{n} -1$ differences and one final sum. 
For the mathematical illustration, we consider a
one-dimensional phase-space, described by the dimensionless variable
$x$ in the interval $[0,1]$. We can divide this phase space in $2^j$
bins of size $\Delta x$ = $1/2^{j}$, where $j$ is a positive integer $j
< j_{max}$, with $j_{max}$  corresponding to the finest resolution
that can be attained. Let us consider  a function $f(x)$
represents any observable on this interval such that
\begin{equation}
\label{mother}
 f^{(j)}(x)=\sum_{k=0}^{2^j-1}f_k^j\,\phi_k^j(x)
\end{equation}
where $\phi_k^j(x)$ is given by
\begin{eqnarray}
 \phi^j_k(x)=\left\{ 
 \begin{array}{ll}
   1 & k/2^j\leq x < (k+1)/2^j \\
   0 & {\rm otherwise}
 \end{array} \right.\,.
\end{eqnarray}

$f_k^j(x)$ is the value of  $f(x)$ in the $k^{th}$ bin.
The family of bin bin functions $\phi^j_k(x)$ can be rewritten as the translations 
and dilation of a single function $\phi(x)$, called the mother function:
\begin{equation}
\label{motherfunc}
 \phi^n_m(x)=\phi(2^nx-m)\,,
\end{equation}
where, in the present case, $\phi(x)$ is the top-hat function.

In Eq.~\ref{motherfunc}, the index $j$ denotes the resolution scale
and $k$ the position at scale $j$. In the multiresolution analysis the
sample  function $f(x)$ can be find at various scales. For example, to
find the structure at lower scale one can replace two adjacent bins
$2k$ and  $2k + 1$ by a single bin of size $2\Delta x = 1/2^{j−1}$ and
corresponding bin function is given by
\begin{equation}
 \phi^{j-1}_k(x)=\phi^j_{2k}(x)+\phi^j_{2k+1}(x)\,,
\end{equation}

In the new bin $k$, by defining the value $f^{j-1}_k$ of the function
as the average of the values in the previous smaller bins:
\begin{equation}
 f^{j-1}_k=\frac{1}{2}(f^j_{2k}+f^j_{2k+1})\,.
\end{equation}
The resulting sample function at scale $j-1$ is given by:
\begin{equation}
\label{mother2}
 f^{(j-1)}(x)=\sum_{k=0}^{2^{j-1}-1}f^{j-1}_k\,\phi^{j-1}_k(x)\,.
\end{equation}

$f^{j-1}_k$'s are called the mother function coefficients (MFCs) and
the Eq.~\ref{mother2}  express the mother function representation of
the distribution $f(x)$ at scale $j-1$. This procedure can be repeated
from the finest resolution scale $j_{max}$ to the lowest one $j=0$.

In the above procedure, another information that can be encoded is the  
difference $\tilde f^{(j-1)}(x)\equiv f^{(j)}(x)-f^{(j-1)}(x)$ and
similarly can be represented as:
\begin{equation}
\label{FFrep}
 \tilde f^{(j-1)}(x)=\sum_{k=0}^{2^{j-1}-1}\tilde f^{j-1}_k\,
 \psi^{j-1}_k(x)\,,
\end{equation}

where the $\psi^{j-1}_k$'s are called father function coefficients
(FFCs) and given by:
\begin{equation}
 \psi^{j-1}_k(x)=\phi^j_{2k}(x)-\phi^j_{2k+1}(x)\,.
\end{equation}
The FFCs can be obtained by the operation of  translations and
dilations $\psi^j_k(x)=\psi(2^jx-k)$.

The FFCs in Eq.~\ref{FFrep} are related to MFCs at previous scale $j$ as
\begin{equation}
\label{FFC}
 \tilde f^{j-1}_k=\frac{1}{2}(f^j_{2k}-f^j_{2k+1})\,.
\end{equation}
The FFCs at a given scale $j$ measure the variation
of the sampled distribution $f$ between two adjacent bins.

To identify jet-like events, we have divided ($\eta - \phi$) region in
both the physical space and scale space. The Haar basis is easy to
visualize but not localized in scale space as the top-hat function is
discontinuous. The simplest wavelet which is localized both in space
and in scale is often called D4-wavelet~\cite{D4_wavelet}. 
Througout our analysis we have used  D4-wavelet to investigate the 
scale dependence of multiplicity fluctuations.

In our analysis the input is the multiplicity in different $\eta$ and
$\phi$  coverages. We have divided the accesible space in 16, $\eta -
\phi$ bins (4 bins in $\eta$ in the range $2.3 < \eta < 3.9$ and 4 bins
in $\phi$ in the range $0 < \phi < 2 \pi$ ) and measure the
multiplicity in each bin.  So in total we have a input column matrix
of $2^{4}$  dimension.  The wavelet transformation give us the bin to 
bin fluctuation, in our case multiplicity fluctuations, in the form of
wavelet coefficients (FFCs). In the next steps it sums up the bins and
again calculates the coefficients and so on till it reaches the step
 where the only difference between last two bins remain.

In Minimum Bias case, we expect the multiplicity distribution over 
many events would be uniform in all the bins which will result in
 no or very less multiplicity fluctuations and hence wavelet 
coefficients approach to a small value. But in case of jet events, 
most of the particle in an event will be localized in a single or two
 bins which will results in higher fluctuations and wavelet
 coefficients attain higher and higher values. At a level or bin 
size in $\eta - \phi$, where the values of the fluctuation 
coefficients are higher gives us an estimate of the size of the jet.

%\leftlinenumbers
\section{Results}
\begin{figure}
     \resizebox{0.35\textwidth}{!}{\includegraphics*{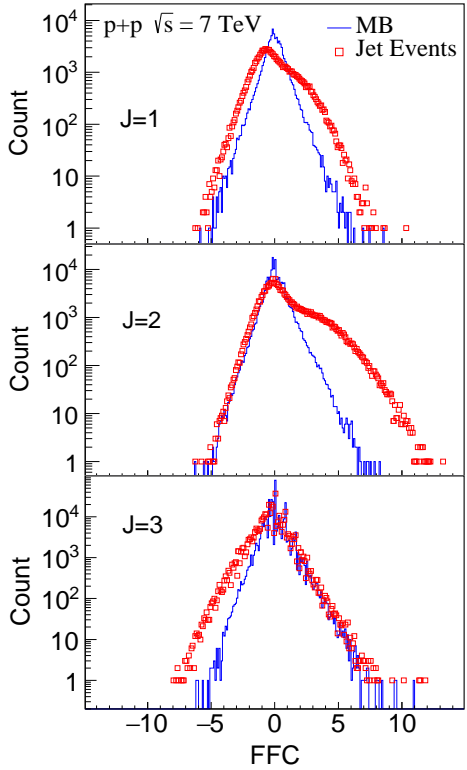}}
     \caption{\label{fig:Chap6_ffcKine}  FFCs distribution for scale J=1, 2, 3 in p-p collisions at $\sqrt{s}$ = 7
  TeV from MC truth. The jet events have jet $E_{T}$ $>$ 20 GeV.}
   \end{figure}

\begin{figure}
\resizebox{0.35\textwidth}{!}{\includegraphics*{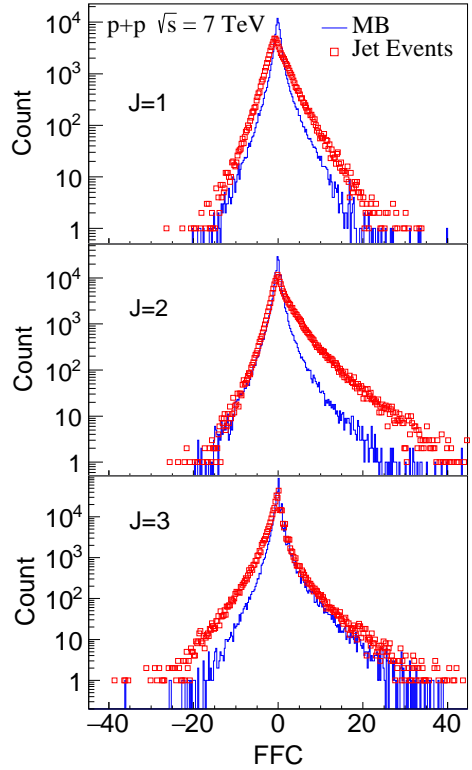}}
\caption{\label{fig:Chap6_ffcDigit} FFCs distribution for scale J=1, 2, 3 in p-p collisions at $\sqrt{s}$ = 7
  TeV from Digits. The jet events have jet $E_{T}$ $>$ 20 GeV.}
\end{figure}

Using wavelet transformations, we obtain the first father
coefficients and plot these coefficients for MB as well
as jet-events and try to quantify the results. In Fig.~\ref{fig:Chap6_ffcKine}
FFC distributions in different scales from MC truth are plotted.
 From the plots,  a clear difference in FFCs distribution for MB and jet
 events are observed. The distribution is broader for jet-events sample as the
 fluctuations are large in these events. Also the broadening of FFCs
 distribution is large for scale J=2. The FFCs at a given scale carry
 information about the degree of fluctuations at higher scales. Due to
 completeness and orthogonality of the DWT basis, there is no
 information loss at any scale. Broader FFC distribution reflects more
 bin-to-bin multiplicity fluctuations, reflecting a jet-like
 behaviour. The scale describes the size of
 the jets on the detector. Fig.~\ref{fig:Chap6_ffcDigit} shows the
FFC distributions in different scales from Digits and the similar
behaviour is observed as in case of MC truth. 
 In order to quantify the analysis, we
 introduced a parameter $\xi$, called as strength parameter.

\subsection{Strength parameter}
In order to quantify the results and obtaining a cut to select jet
like events, we introduce a strength parameter $\xi$. The value of
strength parameter ($\xi$) is given by :

\begin{equation}
\xi = \frac{\sigma^{2}_{jet} - \sigma^{2}_{mb}}{\sigma_{mb}}\\
\end{equation}

where $\sigma_{jet}$ and $\sigma_{mb}$ are widths of the FFCs
 distributions for jets and MB events respectively.

\begin{figure}
     \resizebox{0.4\textwidth}{!}{\includegraphics*{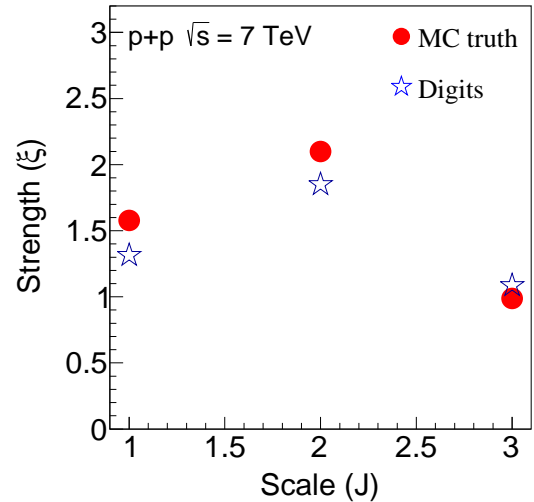}}
\caption{\label{fig:Chap6_chiScale} Strength ($\xi$) vs. scale
 parameter (J) in p-p collisions  at $\sqrt{s}$ = 7 TeV from
 MC truth and Digits.}
   \end{figure}

The parameter $\xi$, simply determines the deviation from normal
distribution.  Figure~\ref{fig:Chap6_chiScale} shows the variation of $\xi$ with scale
parameter J. 
%The left panel is for MC truth and right panel after applying the detector response. 
We can see the variation is similar in both
MC truth and digits level, but the values are little lower at digits
level. This suggests the jet-like signal remains after passing through
detector response. The maximum value of strength parameter is for J=2.

\begin{figure}
     \resizebox{0.39\textwidth}{!}{\includegraphics*{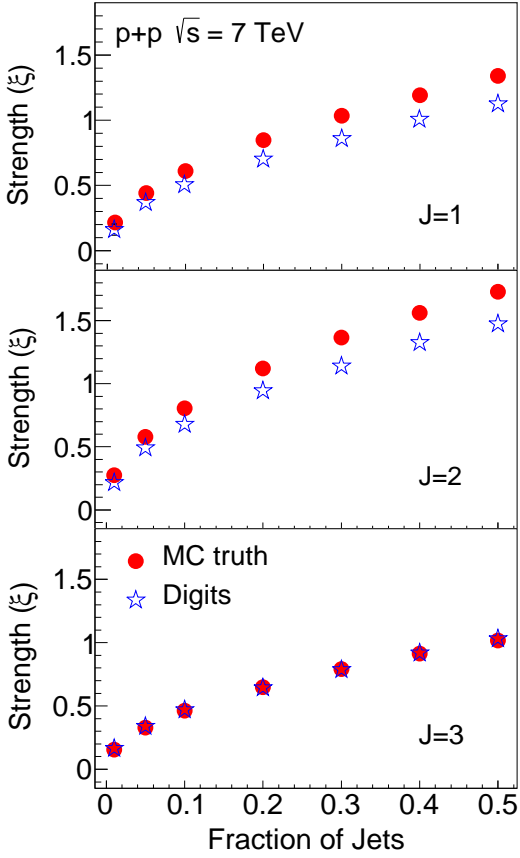}}
\caption{\label{fig:Chap6_chiJJetPercent} Strength ($\xi$) vs. fraction of jet events mixed in MB sample in
   p-p collisions at
   $\sqrt{s}$ = 7 TeV from MC truth and Digits.}
  \end{figure}

\begin{figure}
\resizebox{0.4\textwidth}{!}{\includegraphics*{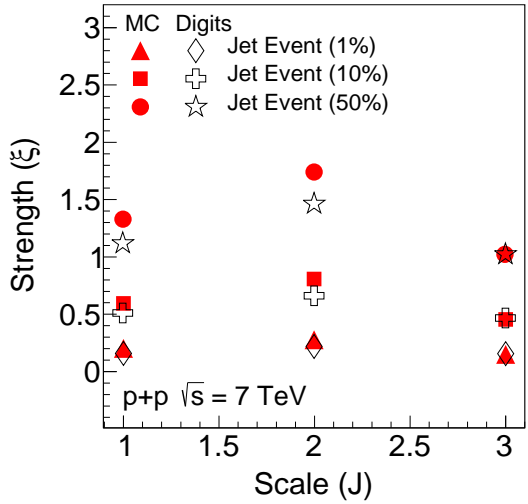}}
\caption{\label{fig:Chap6_chiScaleJetPercent} Strength ($\xi$) vs. scale parameter with different jet
  percentage in p-p collisions at $\sqrt{s}$ = 7 TeV from MC truth and Digits.}
 \end{figure}

The results shown so far are for an ideal situation with full jet
event sample. Now we consider a more realistic scenario of mixing the
MB sample with different percentage of jet-events selected randomly and repeat the same analysis. We have
studied seven such samples with jet event percentage: $1\%$, $5\%$, $10\%$,
$20\%$, $30\%$, $40\%$ and $50\%$. 
%We can see that with increase in the jet event $\%$ the FFCs
%distribution get more and more broadened. 
Strength parameters is
extracted for samples with different jet events percentage and plotted
in Figure~\ref{fig:Chap6_chiJJetPercent}. The figure shows the strength parameter increases with
increasing percentage of jet events. However the magnitude of strength is
higher for MC truth  than Digits. In
Figure~\ref{fig:Chap6_chiScaleJetPercent} strength 
parameter is plotted against the scale
J=1, 2, 3 and the magnitude is higher for J=2 both for MC truth and Digits.

\subsection{Model comparison}
\begin{figure}
 \resizebox{0.39\textwidth}{!}{\includegraphics*{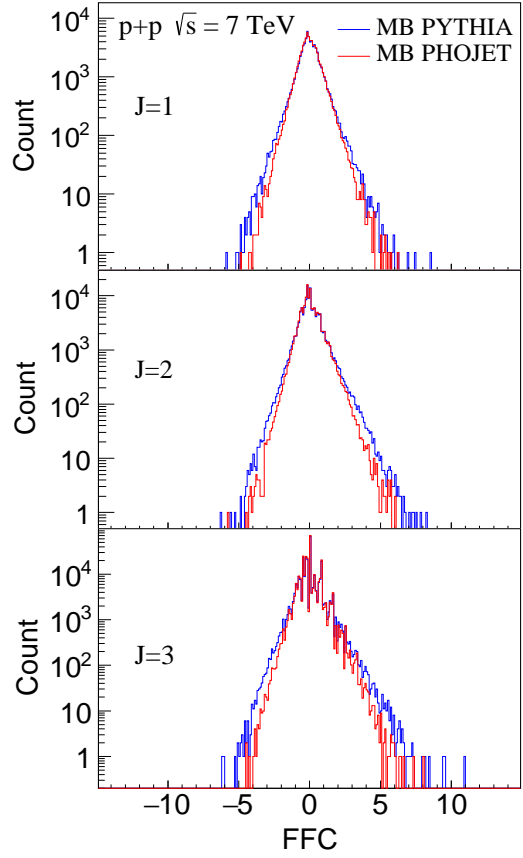}}
\caption{\label{fig:Chap6_ffcModel} FFCs distribution in PYTHIA6 and PHOJET for scale J =
   1, 2, 3 in p-p collisions at $\sqrt{s}$ = 7 TeV for MC truth.}
\end{figure}

It is important here to check for any model dependence of present
analysis. For this we have carried out the analysis with another event
generator PHOJET \cite{phojet}. The FFCs distribution for events
generated from PHOJET are compared to those from PYTHIA6 \cite{pythia}. Figure~\ref{fig:Chap6_ffcModel} shows 
the FFCs distribution for both the models for different scales.
From the figure it is clear that there is very good agreement in the
models and hence our results are model independent.

%%%% I am HERE for model comparison

\subsection{Energy dependence}
\begin{figure}
 \resizebox{0.4\textwidth}{!}{\includegraphics*{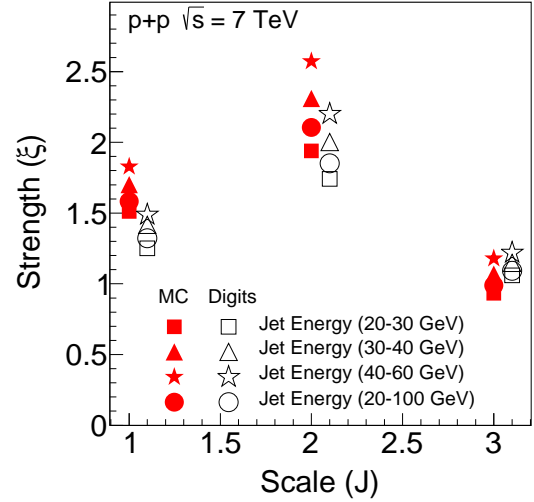}}
\caption{\label{fig:Chap6_energyDep} Strength ($\xi$) vs. scale
 parameter for different jet transverse energies in p-p collisions at
 $\sqrt{s}$ = 7 TeV from MC truth and Digits.}
\end{figure}

Figure~\ref{fig:Chap6_energyDep} shows the strength parameter dependence on jet transverse
energy for scale J=1, 2, 3 for MC truth 
and Digits. We have generated four events
sample with jet $E_{T}$ ranging; 20-30 GeV, 30-40 GeV, 40-60 GeV and
20-100 GeV. From the figure we can see that the magnitude of $\xi$ is
highest for jet $E_{T}$ = 40-60 GeV i.e. for high jet $E_{T}$ case.\\

\subsection{Efficiency, purity of selecting jet like events}
\begin{figure}
 \resizebox{0.4\textwidth}{!}{\includegraphics*{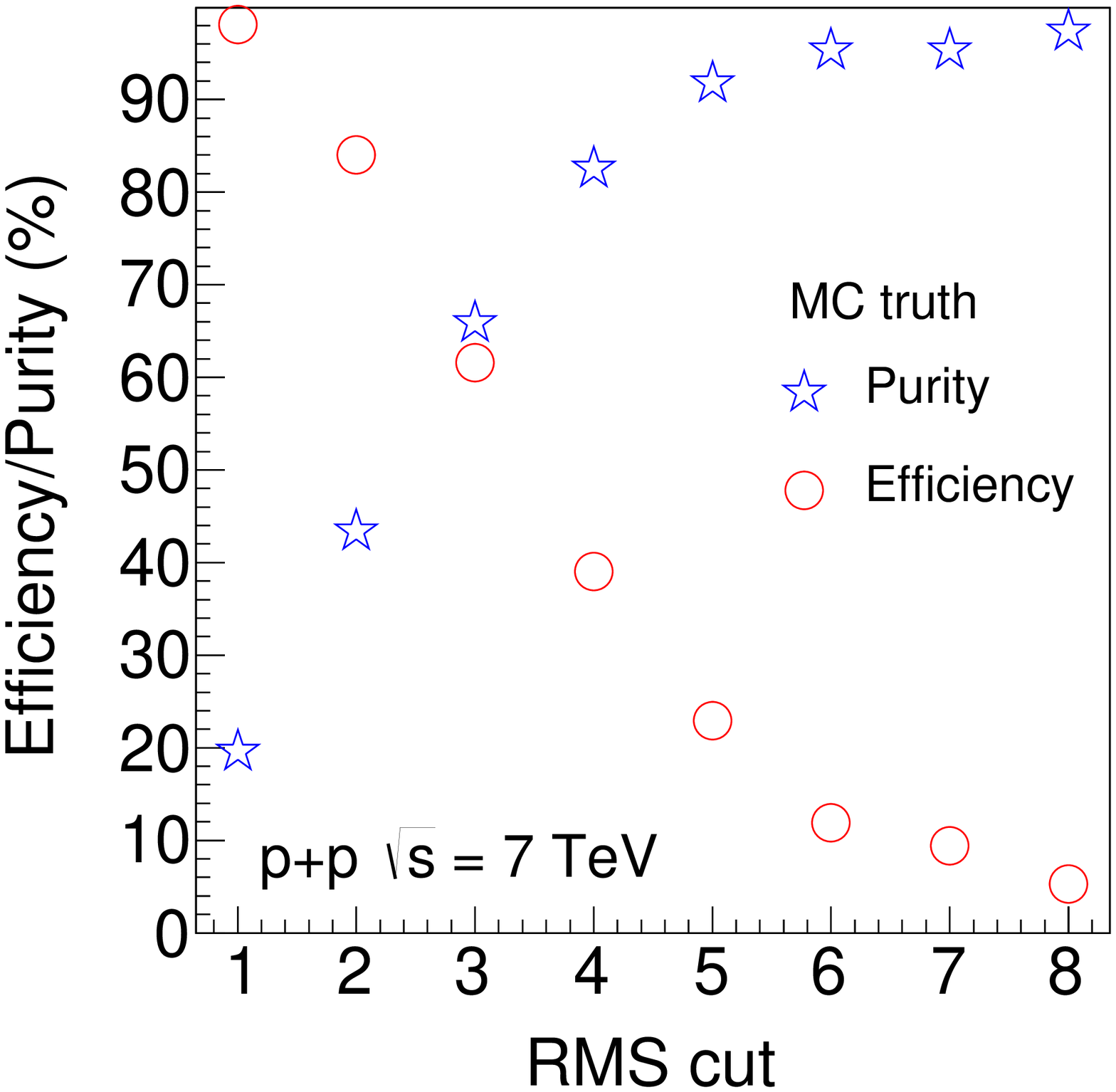}}
\caption{\label{fig:Chap6_effy} Efficiency, Purity vs. rms cut on MB
  FFCs distribution at
  $\sqrt{s}$ = 7 TeV  from MC truth.}

\resizebox{0.4\textwidth}{!}{\includegraphics*{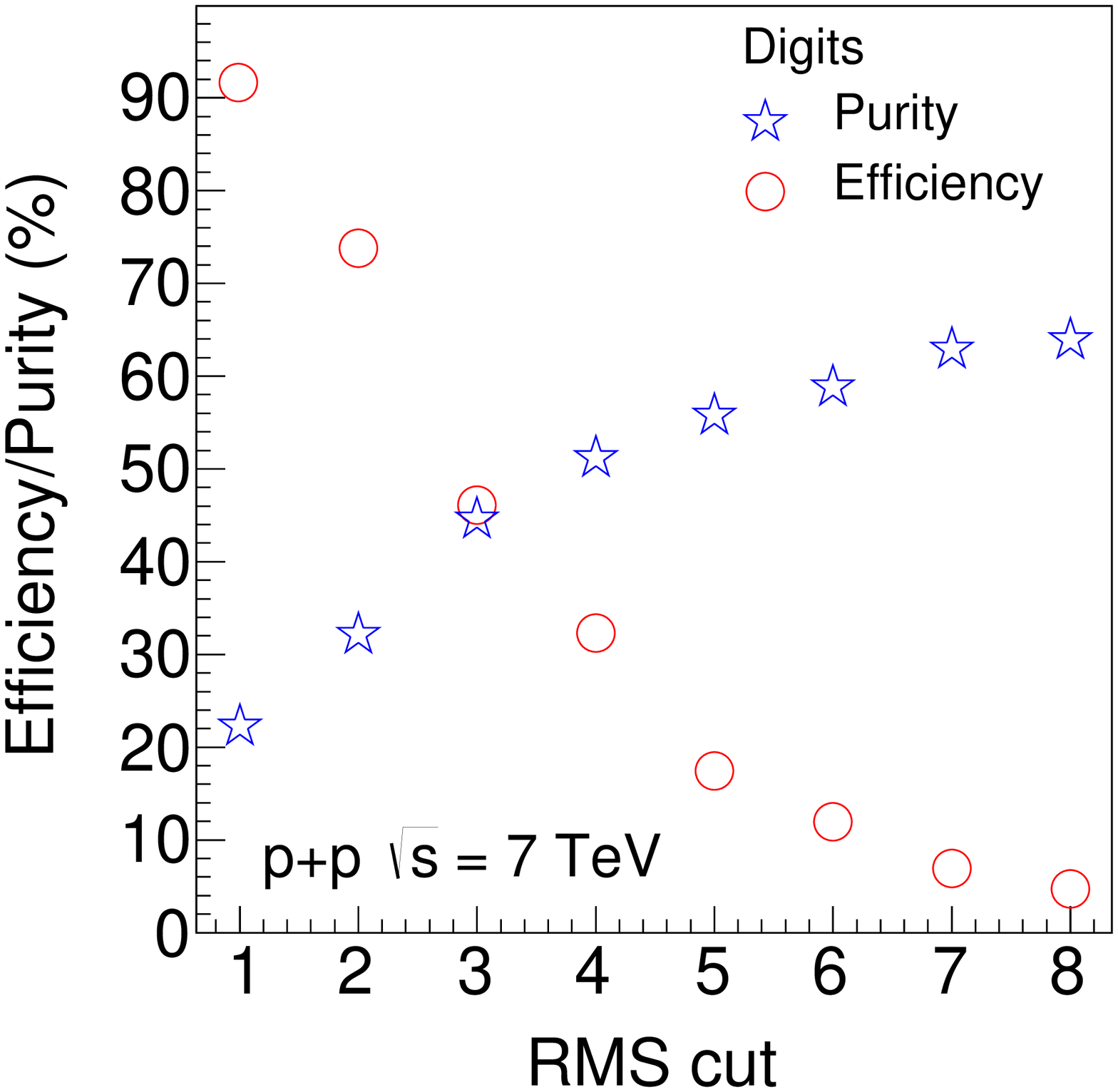}}
\caption{\label{fig:Chap6_purity} Efficiency, Purity vs. rms cut on MB
  FFCs distribution at
  $\sqrt{s}$ = 7 TeV  from Digits.}
\end{figure}

Efficiency and purity of selecting jet-like events in a multiplicity
detector is calculated by applying different rms cuts
from normal FFCs distribution on a sample of mixed MB+jet events with $50\%$
jets. The efficiency and purity are defined as :

\begin{equation}
%\begin{center}
Efficiency = \frac{N_{j}}{N_{ja}}
%\end{center}
\end{equation}

\begin{equation}
%\begin{center}
Purity = \frac{N_{j}}{N_{jlike}}
%\end{center}
\end{equation}

where, $N_{j}$ = No. of jet events identified, $N_{ja}$ = No. of jet events added
in the MB sample and $N_{jlike}$ = No. of jet like events.
Figures~\ref{fig:Chap6_effy} $\&$~\ref{fig:Chap6_purity} show the 
variation of efficiency / purity with the rms
cuts obtained from the FFCs distribution of MB events for MC truth
 and Digits respectively. For a cut of 3 times the rms of FFCs
 distribution of MB events the efficiency and purity
 values are about 46$\%$.

\section{Summary} 
\noindent 
We have reported a multi-resolution analysis technique to identify
jet-like events in a multiplicity detector at LHC
energies. The analysis is carried out using charged particle
multiplicities. The observation of jet-like events can be used to tag 3-jet events
in ALICE at LHC. The multi-resolution simulation study shows the good
sensitivity towards selecting jet-like events in the forward
multiplicity detector. The
value of $\xi$ is 0.22, 0.65 \& 1.42 for event samples with 1\%, 10\%
and 50\% jet-like events respectively. A $\xi$ value of zero would
have indicated no sensitivity of the method towards identification of
jet-like events. Such an analysis can be carried out in real data in
future with p-p collisions at 0.9, 2.76, 7, 8 and 13TeV data collected by
ALICE.

 \section*{Acknowledgement}
 \noindent We would like to thank Y. P. Viyogi for many helpful
 discussions on tagging jet-like events using a multiplicity
 detector. The authors acknowledge the  support from DST SwarnaJayanti
and DAE-SRC projects of Govt. of India.

\end{document}